\documentclass[12pt]{article}
\usepackage{amsmath}
\usepackage{graphicx}
\usepackage{cite}
\usepackage{url}
\usepackage{xcolor}
\usepackage[colorlinks,citecolor=blue,urlcolor=blue,linkcolor=blue]{hyperref}

\newcommand\pubnumber{}
\newcommand\pubdate{\today}

\def\institute{Physics Department, Florida State University \\
Tallahassee, FL 32306-4350, U.S.A}
\def\support{\footnote{Work is supported in part by U.S. Department of Energy under grant DE-SC010102}}

\def\Title#1{\begin{center} {\Large #1 } \end{center}}
\def\Author#1{\begin{center}{ \sc #1} \end{center}}
\def\Address#1{\begin{center}{ \it #1} \end{center}}

\newcommand\pubblock{\rightline{\begin{tabular}{l} \pubnumber\\
         \pubdate  \end{tabular}}}
\newenvironment{Abstract}{\begin{quotation}  }{\end{quotation}}
\newenvironment{Presented}{\begin{quotation} \begin{center} 
             PRESENTED AT\end{center}\bigskip 
      \begin{center}\begin{large}}{\end{large}\end{center} \end{quotation}}

\def\beq{\begin{equation}}
\def\eeq#1{\label{#1}\end{equation}}
\def\eeqn{\end{equation}}

\def\beqa{\begin{eqnarray}}
\def\eeqa#1{\label{#1}\end{eqnarray}}
\def\eeqan{\end{eqnarray}}

\let\bar=\overbar

\def\Dslash{\not{\hbox{\kern-4pt $D$}}}
\def\dslash{\not{\hbox{\kern-2pt $\del$}}}

\def\msb{{\bar{\ssstyle M \kern -1pt S}}}

\def\helacnlo{\textsc{Helac-NLO}}
\def\helloop{\textsc{Helac-1Loop}}
\def\heldip{\textsc{Helac-Dipoles}}

\begin{document}
\begin{titlepage}
\pubblock

\vfill
\Title{off-shell $t\bar{t}b\bar{b}$ in the di-lepton channel}
\vfill
\Author{ Manfred Kraus\support}
\Address{\institute}
\vfill
\begin{Abstract}
We report on our recent calculation for the off-shell $t\bar{t}b\bar{b}$
process in the di-lepton decay channel at the LHC. Our results take into
account NLO QCD corrections for the complete $pp\to e^+\nu_e\mu^-
\bar{\nu}_\mu b\bar{b}b\bar{b}$ process, and include all double, single and
non-resonant contributions. We investigate the size of the corrections and
their associated theoretical uncertainties.  We also briefly comment on the
impact of different $b$ jet definitions on our results.
\end{Abstract}
\vfill
\begin{Presented}
$14^\mathrm{th}$ International Workshop on Top Quark Physics\\
(videoconference), 13--17 September, 2021
\end{Presented}
\vfill
\end{titlepage}
\def\thefootnote{\fnsymbol{footnote}}
\setcounter{footnote}{0}
%

\section{Introduction}
With the discovery of the Higgs boson at the LHC the focus of the physics
programs lies on the precise investigation of fundamental properties of the
Higgs boson and its interactions. In that context the top-quark Yuakawa
coupling is of special interest as the top-quark is the heaviest fundamental
particle in the Standard Model. To probe directly the top-higgs interactions it
is beneficial to study the Higgs boson production in association with a
top-quark pair. In order to compensate for the small production rate of the $pp
\to t\bar{t}H$ process the decay of the Higgs boson into bottom quarks, $H\to
b\bar{b}$, is of special interest as it has the largest branching ratio.

To measure precisely the $pp \to t\bar{t}H(H\to b\bar{b})$ signal multiple
challenges have to be overcome. For instance, once also the top-quarks decay
the final state consists out of four $b$ jets, which generates the
combinatorical problem of the $b$ jet assignment in the reconstruction of the
Higgs boson. On the other hand, large Standard Model background processes have
to be under excellent theoretical control to isolate signal events. The main
background contributions consists out of the QCD production of the $pp\to
t\bar{t}b\bar{b}$ process at $\mathcal{O}(\alpha_s^4)$.
Similarly, due to the presence of the four $b$ jets in the decayed signature,
$pp \to t\bar{t}b\bar{b}$ also affects searches for the four top-quark
production at the LHC. Thus, measurements of $pp\to t\bar{t}H(H\to b\bar{b})$
and $t\bar{t}t\bar{t}$ will benefit from a better understanding of the QCD
production of the $pp\to t\bar{t}b\bar{b}$ process as well as the improved
description of top-quark decays.

The $t\bar{t}b\bar{b}$ process has been extensively studied in the literature
at NLO QCD accuracy. For instance, for on-shell $t\bar{t}b\bar{b}$ the
next-to-leading order QCD corrections are
available~\cite{Bredenstein:2008zb,Bredenstein:2009aj,
Bevilacqua:2009zn,Bredenstein:2010rs,Worek:2011rd} for over $10$ years by now.
The impact of parton shower matching has been also thoroughly addressed in
Refs.~\cite{Kardos:2013vxa,Cascioli:2013era,Garzelli:2014aba,Bevilacqua:2017cru,Jezo:2018yaf}.
Furthermore, $t\bar{t}b\bar{b}/t\bar{t}jj$ cross section ratios have been
investigated in Ref.~\cite{Bevilacqua:2014qfa} as well as the
$t\bar{t}b\bar{b}$ production in association with a hard
jet~\cite{Buccioni:2019plc} has been computed at NLO QCD accuracy.  In all the
aforementioned studies top-quarks have been treated as stable particles and
top-quark decays have been included at most at LO accuracy.  Only recently the
computation of NLO QCD corrections for the full off-shell process in the
di-lepton channel~\cite{Bevilacqua:2021cit,Denner:2020orv} has become feasible.
These calculations include for the very first time also NLO QCD corrections to
the top-quark decays and double, single and non-resonant contribution as well
as interference effects.  Here we report on our recent calculation for the full
off-shell production of the $pp\to t\bar{t}b\bar{b}$
process~\cite{Bevilacqua:2021cit} including leptonic top-quark decays. 

\section{Outline of the calculation}
\begin{figure}[h!]
\begin{center}
  \includegraphics[width=\textwidth]{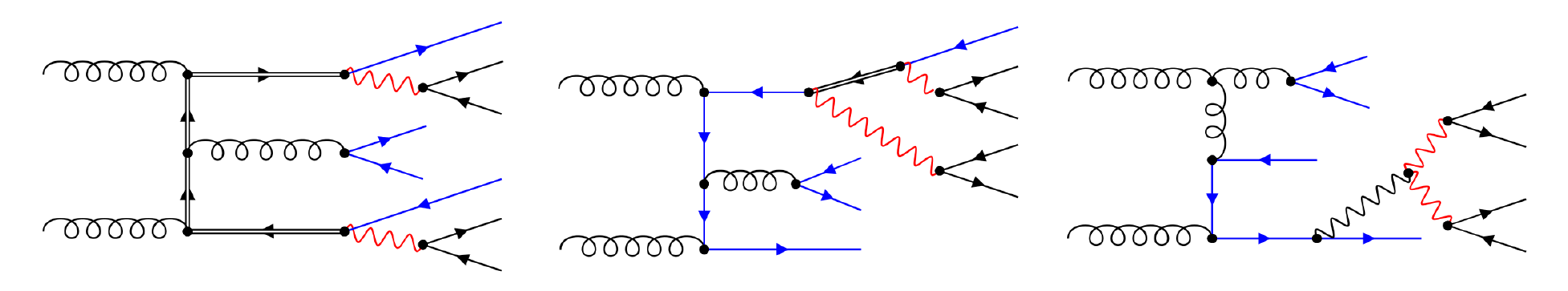}
\end{center}
\caption{Representative Feynman diagrams for double, single and non-resonant
contributions in the $pp\to b\bar{b}b\bar{b}e^+\nu_e\mu^- \bar{\nu}_\mu$ matrix
elements. Figure taken from Ref.~\cite{Bevilacqua:2021cit}}
\label{fig:diags}
\end{figure}

In the following we will briefly summarize the outline of the calculation.  For
more details we refer the reader to the original
publication~\cite{Bevilacqua:2021cit}.  We compute NLO QCD corrections for the
$pp\to e^+\nu_e\mu^- \bar{\nu}_\mu b\bar{b}b\bar{b}$ process at
$\mathcal{O}(\alpha^4\alpha_s^5)$ for the LHC operating at a center-of-mass
energy of $\sqrt{s}=13$ TeV.  The computation is based on matrix elements for
the fully decayed final state that comprise not only the double resonant
$t\bar{t}b\bar{b}$ production, but also single and non-resonant contributions,
for which representative diagrams are shown in Fig.~\ref{fig:diags}.
In addition, the matrix elements also account for finite width effects of
unstable particles and interference effects between the various contributions.
The calculation is performed using the \helacnlo{}
framework~\cite{Bevilacqua:2011xh} that has been already employed in various
full off-shell computations at NLO QCD accuracy for $t\bar{t}V$
processes~\cite{Bevilacqua:2015qha,Bevilacqua:2016jfk,Bevilacqua:2018woc,Bevilacqua:2019cvp,Bevilacqua:2020pzy}.
The framework consists out of
\helloop{}~\cite{Ossola:2007ax,vanHameren:2009dr,vanHameren:2010cp} for the
computation of virtual one-loop corrections and
\heldip{}~\cite{Czakon:2009ss,Bevilacqua:2013iha,Czakon:2015cla} that takes
care of the infrared subtraction for radiative corrections. 

We investigate a fiducial signature consisting out of at least four $b$ jets,
two charged leptons and missing energy. Jets are formed using the anti-$k_T$
jet algorithm~\cite{Cacciari:2008gp} with a separation parameter of $R=0.4$.
We employ the following inclusive phase space cuts
\begin{equation}
 p_T(\ell) > 20~\text{GeV}\;, \quad p_T(b) > 25~\text{GeV}\;, \quad |y(\ell)| < 2.5\;, \quad |y(b)| < 2.5\;.
\end{equation}
%
\section{Phenmenological results}
Let us start our discussion with the inclusive cross section. In
Tab.~\ref{tab:xsec} LO and NLO integrated cross sections are shown for two
different choices of renormalization and factorization scales
$\mu_R=\mu_F=\mu_0$, namely a fixed scale $\mu_0 = m_t$ and a dynamical one
$\mu_0 = H_T/3$. Also shown are scale uncertainties estimated from independent
variations of the renormalization and factorization scales as well as PDF
uncertainties.
\begin{table}[h!]
\begin{center}
\begin{tabular}{ccccccc}
 \hline \hline\\[-0.4cm]
 $\mu_0$ & $\sigma^{\rm LO}$ [fb] & $\delta_{\rm scale}$ &  $\sigma^{\rm NLO}$ [fb] & $\delta_{\rm scale}$ & $\delta_{\rm PDF}$  &  ${\cal K}=\sigma^{\rm NLO}/\sigma^{\rm LO}$ \\[0.2cm]
 \hline\hline\\[-0.4cm]
 $m_t$ & $6.998$ & ${}^{+4.525~(65\%)}_{-2.569~(37\%)}$ & $13.24$ & ${}^{+2.33~(18\%)}_{-2.89~(22\%)}$ & ${}^{+0.19~(1\%)}_{-0.19~(1\%)}$ & $1.89$\\[0.2cm] 
 \hline\hline\\[-0.4cm]
 $H_T/3$ & $6.813$ & ${}^{+4.338~(64\%)}_{-2.481~(36\%)}$ & $13.22$ & ${}^{+2.66~(20\%)}_{-2.95~(22\%)}$  & ${}^{+0.19~(1\%)}_{-0.19~(1\%)}$ & $1.94$\\[0.2cm] 
 \hline \hline   
\end{tabular}
\end{center}
\caption{\label{tab:xsec} 
  LO and NLO integrated fiducial cross sections for the $pp\to
e^+\nu_e\, \mu^-\bar{\nu}_\mu\, b\bar{b} \,b\bar{b}$ process at the
LHC with $\sqrt{s}=13$ TeV.}
\end{table}
We find that the NLO QCD corrections are large and of the order $+90\%$. At the
same time the residual scale dependence reduces roughly by a factor of $3$ from
$60\%$ at LO down to the level of $20\%$ at NLO. Given the still sizable scale
uncertainties at NLO the PDF uncertainties, which are of the order of $\pm 1\%$
are negligible. Our findings are the same independent of the nature of the
employed renormalization and factorization scale. However, for differential
distributions we only show results for the dynamical scale $\mu_0 = H_T/3$, as
it is well known that a fixed scale can not capture the behaviour of the
high-energy tails appropriately.

\begin{figure}[h!]
\begin{center}
  \includegraphics[width=0.48\textwidth]{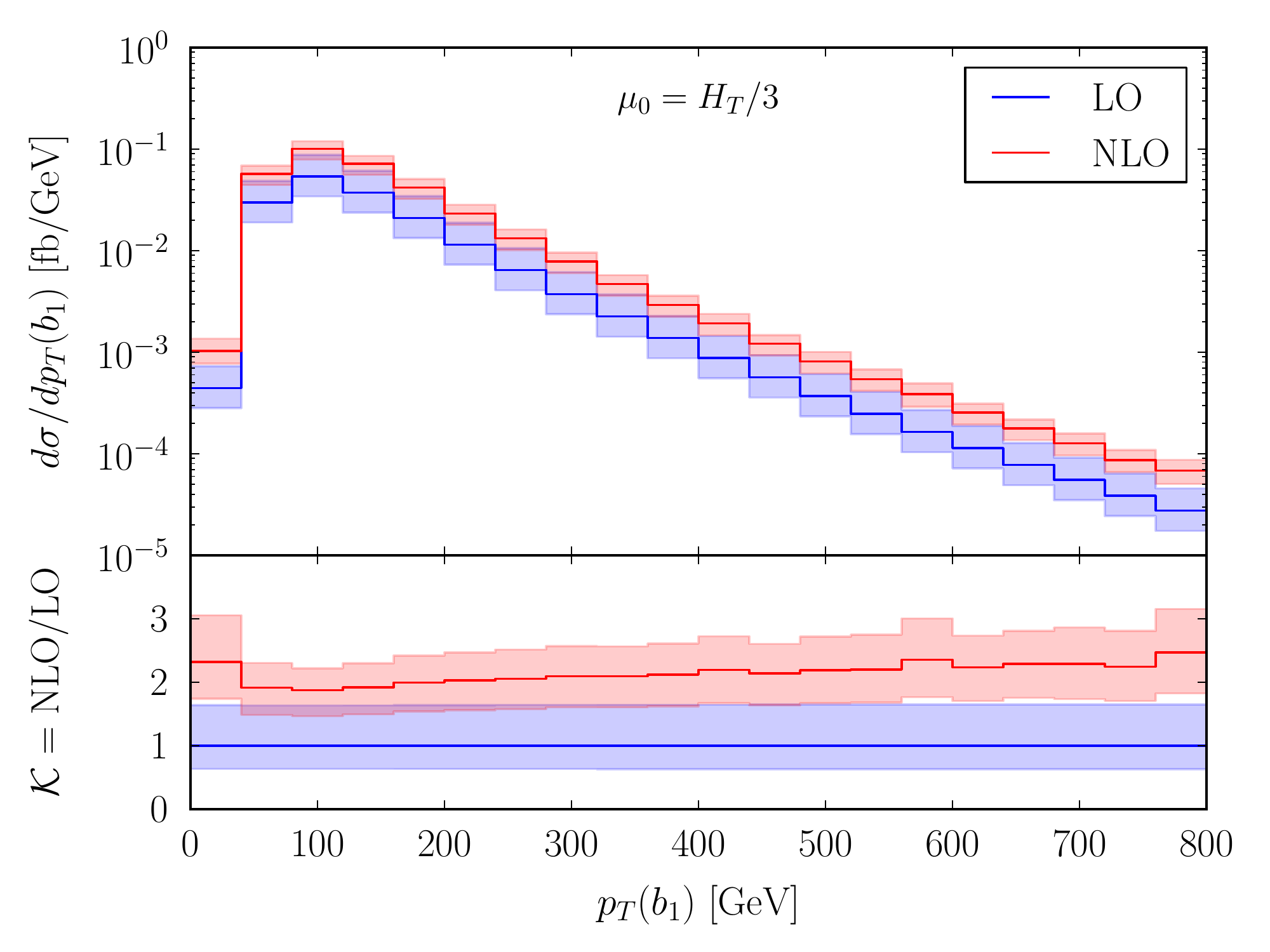}
  \includegraphics[width=0.48\textwidth]{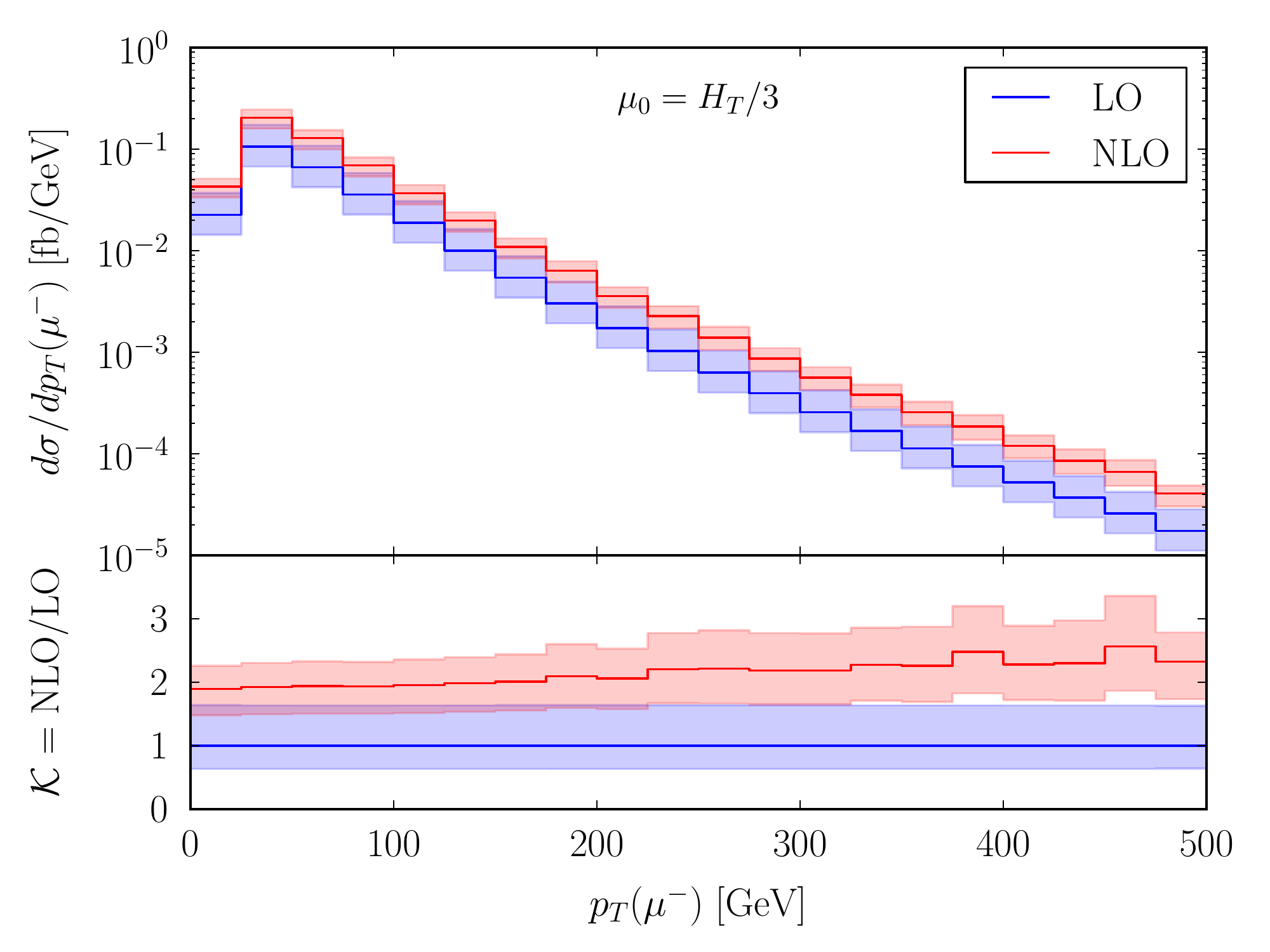}
\end{center}
\caption{Differential cross section distribution for the transverse momentum 
of the hardest $b$ jet (l.h.s) and of the muon (r.h.s). 
Plots taken from Ref.~\cite{Bevilacqua:2021cit}.}
\label{fig:diff}
\end{figure}

Next we discuss the impact of NLO QCD corrections at the differential level.
In Fig.~\ref{fig:diff} we show the differential cross section distribution of
the transverse momentum of the hardest $b$ jet and of the muon.
Even though the transverse momentum of the hardest $b$ jet is a hadronic
observable, whereas the transverse momentum of the muon is a purely leptonic
one they are affected very similar from NLO QCD corrections.  The corrections
are large over the whole plotted range and increase in the tail of the
distribution. Thus the NLO K-factor is far from being flat. Furthermore, we
notice that the scale uncertainty bands are barely overlapping and have similar
sizes suggesting a strong impact from real radiation processes that are only
taken into account at LO accuracy. 

We have also investigated PDF uncertainties at the differential level, which 
are at the level of a few percent for most observables. However, in extreme
cases they can be as large as $\pm 10\%$. We refer the reader to our 
publication~\cite{Bevilacqua:2021cit} for more details.

At last, we also estimated the impact of different definitions of $b$ jets on
our results. We consider the two cases where one can either distinguish between
$b$ and $\bar{b}$ or not.  This leads to the following recombination rules for
the jet algorithm
\begin{equation}
\begin{split}
 \text{charge-blind:}& \qquad bg\to b\;, \bar{b}g\to \bar{b}\;, b\bar{b}\to g\;, bb\to g\;, \bar{b}\bar{b}\to g\;, \\
 \text{charge-aware:}& \qquad bg\to b\;, \bar{b}g\to \bar{b}\;, b\bar{b}\to g\;, bb\to b\;, \bar{b}\bar{b}\to \bar{b}\;.
\end{split}
\end{equation}
Notice that in the charge-aware scheme the cross section does not receive
contributions from initial state $bb$ and $\bar{b}\bar{b}$ processes, due to
the requirement that there are at least $2$ b jets and $2$ anti-$b$ jets in the
final state. At the level of inclusive cross sections we find differences of
the order of $1\%$ with respect to simply ignoring initial state $b$
contributions.
\begin{figure}[h!]
\begin{center}
  \includegraphics[width=0.48\textwidth]{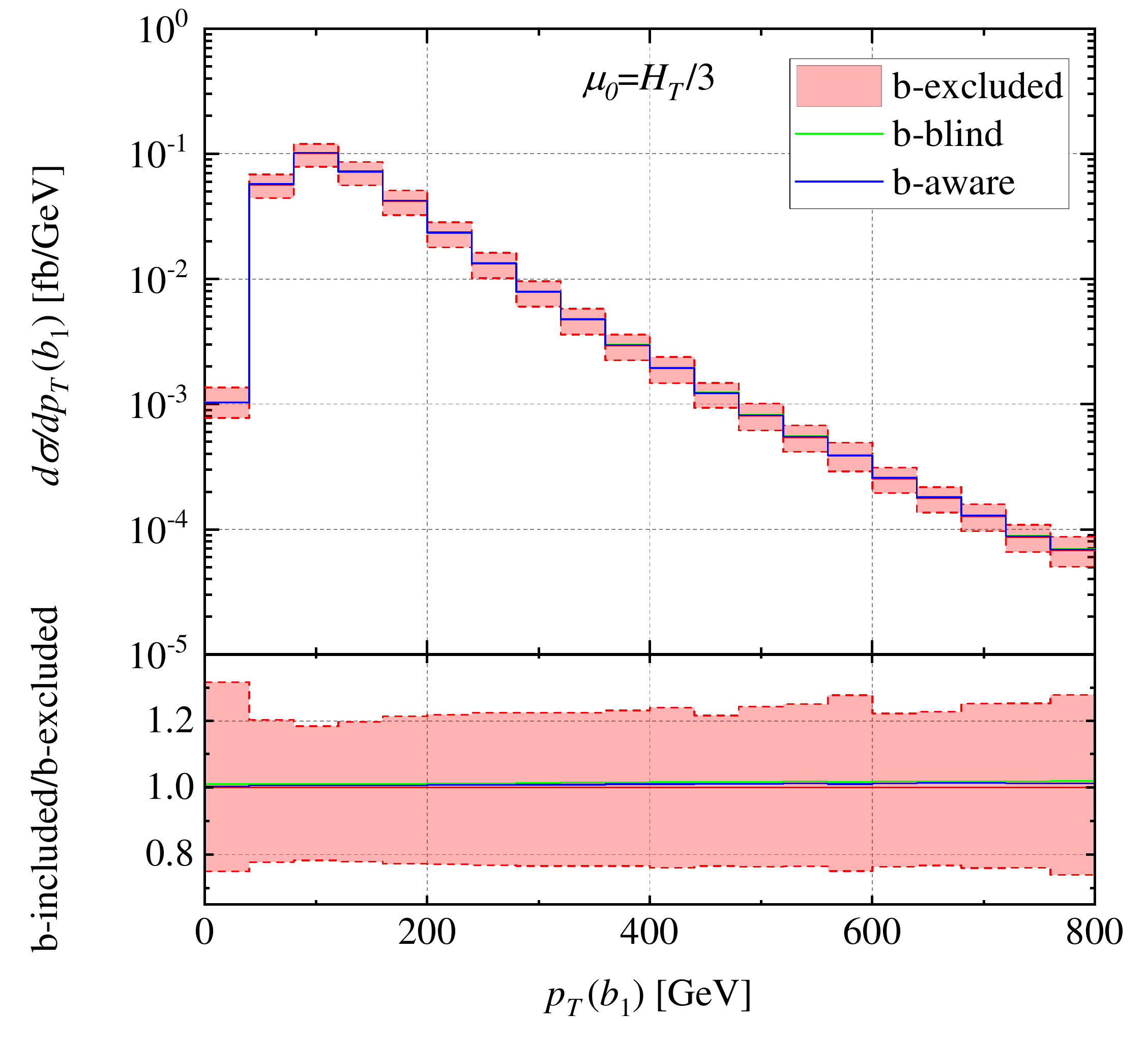}
  \includegraphics[width=0.48\textwidth]{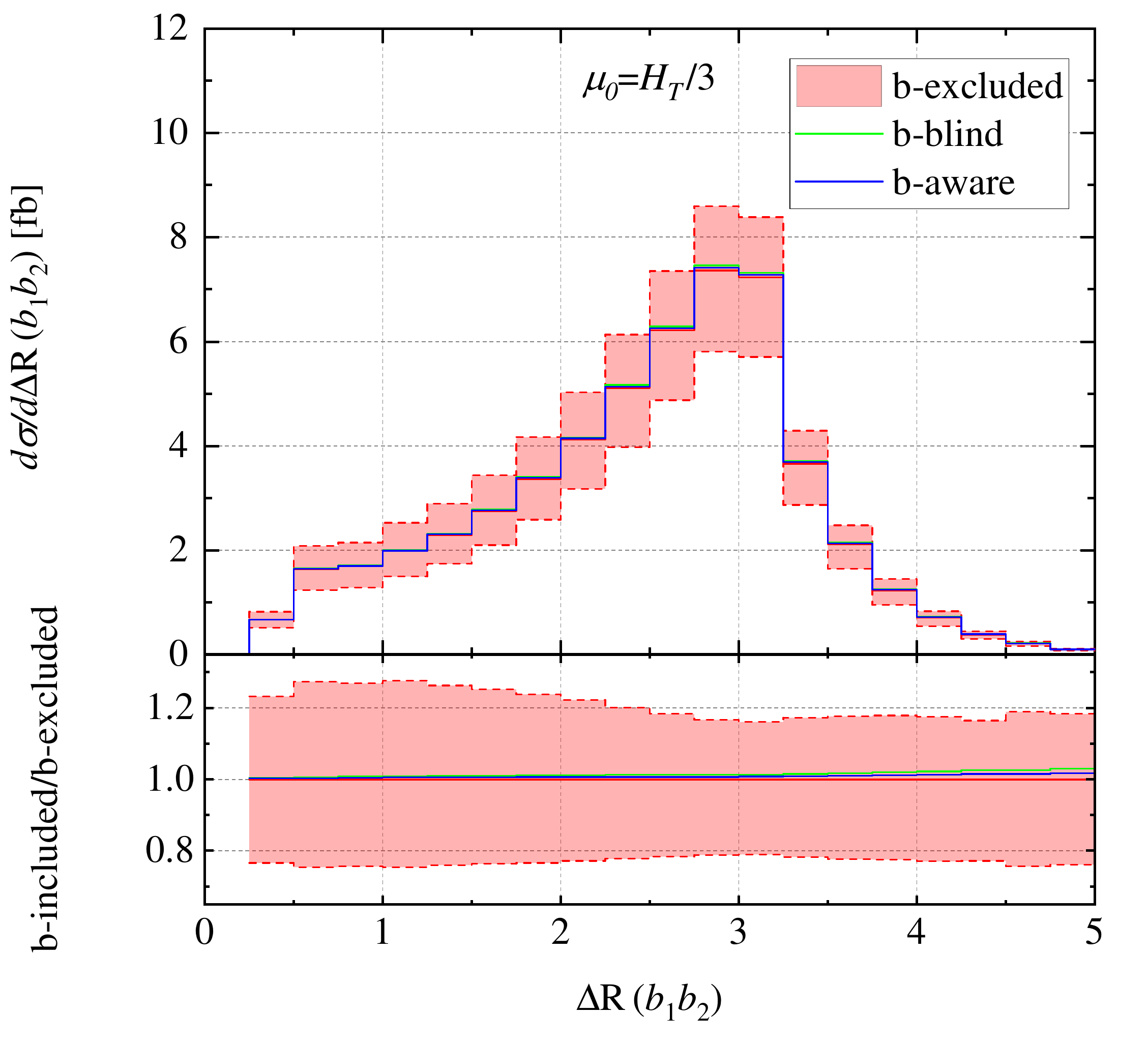}
\end{center}
\caption{Differential cross section distribution for the transverse momentum of
the hardest $b$ jet (l.h.s) and the $\Delta R$ separation of the two hardest
$b$ jets (r.h.s) for various $b$ jet definitions.  Plots taken from
Ref.~\cite{Bevilacqua:2021cit}.}
\label{fig:diff_b}
\end{figure}
However, at the differential level these could be sizable in certain phase
space regions.  In Fig.~\ref{fig:diff_b} we show the impact of the various $b$
jet definitions for the case of the transverse momentum of the hardest $b$ jet
as well as the $\Delta R$ separation between the two hardest $b$ jets. Also at
the differential level we observe that initial state $b$ contributions are
negligible. Even in extreme phase space regions, for example $\Delta R \gg 3$,
which could in principle be sensitive to the $gb$ induced real radiation
processes do not show a significant enhancement. Therefore, initial state
$b$ contributions are generally deemed negligible.

\section{Conclusions}
We highlighted some of the recent results~\cite{Bevilacqua:2021cit} of our
state-of-the-art calculation for off-shell $pp\to t\bar{t}b\bar{b}$ production
in the di-leptonic decay channel for the LHC at $\sqrt{s}=13$ TeV.  We
discussed briefly the outline of our calculation and the impact of NLO QCD
corrections at the inclusive and differential level.  We found large
corrections of the order of $+90\%$ for the integrated cross section, while at
the differential level they increase even further. The scale uncertainties are
at the $\pm 20\%$ level, while PDF uncertainties, which can amount up to a few
percent are negligible in comparison. In addition, we also investigated the
impact of different $b$ jet definitions on our results. We found that our
results are very stable with respect to modifications of the $b$ jet definition
and that initial state $b$ contributions are overall negligible.

\bibliography{eprint}
\bibliographystyle{JHEP}
 
\end{document}